\documentstyle[11pt]{article}
\begin{document}
\title{{\bf Meson decay in an independent quark model}}
\author{{\bf Hakan \d{C}\.{I}FT\d{C}\.{I}\thanks{
            e--mail: hciftci@quark.fef.gazi.edu.tr}} and
         {\bf H\"{u}seyin KORU\thanks{
            e--mail: hkoru@quark.fef.gazi.edu.tr}} \\
{\it Gazi University, Faculty of Art and Science,
                                06500 Ankara, TURKEY}}
\maketitle

\begin{abstract}
Leptonic decay widths and leptonic decay constants of light
vector mesons and weak leptonic decay widths and weak decay
constants of light and heavy pseudoscalar mesons have been
studied in a field- theoretic framework based on the independent
quark model with a scalar- vector power-law potential.
The results are in very good agreement with the experimental data.
\end{abstract}
\vspace{1cm}
~~~~PACS numbers: 12.39.Ki, 12.39.Pn,  13.20.-v

\baselineskip 0.8cm

\section{Introduction}

Though quantum chromodynamics is considered to be the underlying
theory of strong interaction between quarks and gluons at the
structural level of hadrons, many low-energy phenomena such as
spectroscopy, static electromagnetic properties, can not be
explained by first-principles application of QCD. Therefore
one needs to resort to phenomenological models. For example,
a potential model with an equally mixed scalar-vector harmonic
potential$^1$ of independent quarks in a relativistic Dirac
framework has been used to study several low-energy phenomena
in the baryonic sector such as octet baryon masses$^{2}$, magnetic
moments$^{3}$, weak electric form factors$^{4}$, nucleon
electromagnetic form factors and charge radii$^{5}$. In addition
to the harmonic potential, with power-law potential such as
$r^\nu $($\nu >0$) heavy-heavy and light-heavy quarkonium states
have been studied and the results are in good agreement with
experimental results$^{6}$. This model has also been successful
in explaining pion mass, its decay constant$^{7}$ and the radiative
decay$^{8}$ of ordinary light and heavy mesons. Because of this
wide range of applicability of the model to both baryons and mesons,
it has proved to be a rather simple and successful alternative
to the cloudy bag model$^{9}$. The purpose of this work is to
extend its applicability to the study of the leptonic decay of
vector mesons such as $\rho $, $\omega $, $\phi $ and weak leptonic
decay of light and heavy pseudoscalar mesons for an equally mixed
scalar-vector power-law potential such as $Ar^\nu$. The leptonic
decay width of heavier vector mesons in the charm and bottom quark
sector have been studied in the nonrelativistic approach through
the Van Royen-Weisskopf formula with radiative corrections$^{10}$.
But the same approach is not suitable for ordinary vector mesons
in the light flavor sector, where the constituent quark dynamics
is more relativistic. On the other hand, for weak leptonic decays,
while many nonrelativistic quark model calculations$^{11}$ suggest
that $f_K>f_D>f_B$ ($f_M$ is the weak leptonic decay constant),
some of the models based on QCD sum rules$^{12}$ and lattice
calculations$^{13}$ predict more or less a constant $f_M$ between
$K$ and $B$ mesons. Capstick and Godfrey$^{14}$ calculate the
hadronic matrix elements for the relativitized quark model
expression for $f_M$. They find
$ f_{B_c}>f_{D_s}>f_K>f_D>f_{B_s}>f_B>f_\pi$, but the calculated
value of the ratio $f_K/f_\pi \cong 1.75$ is much higher than the
experimental value of $1.22$. The leptonic decay widths and decay
constants $f_V$ of the light vector mesons have been calculated
by using an equally mixed scalar vector harmonic potential$^{15}$.
Also, weak leptonic decay constants, $f_M$, of pseudoscalar mesons
have been calculated using the same potential and are found to
satisfy $f_{B_c}>f_{D_s}>f_D>f_K>f_{B_s}>f_\pi >f_B$$^{16}$.
Potential used in Ref. 15 and 16 is harmonic. Since the potential
between quarks is weaker than the harmonic potential, the potential
used in this study is $Ar^{0.2}+V_0$ which gives very good results
for heavy-heavy and light-heavy quarkonium states and radiative
decay of ordinary light and heavy mesons$^{6,8}$. Both in Ref.
15,16 and in this study, the potential does not include Coulomb
term because Coulomb-like vector interaction are believed to have
less prominent role for the light mesons.

\section{Potential model}

The quark-confining interaction in a hadron, which is believed to
be generated by the nonperturbative multigluon mechanism, is not
possible to calculate theoretically from first-principles within QCD.
Therefore, from a phenomenological point of view, the present model
assumes that the quark and antiquark in a hadron core are
independently confined by an average flavor-independent potential
$V(r)$ of the form
\begin{equation}
V(r)=\frac 12(1+\beta )(Ar^\nu +V_0),~~~~A>0~~~\mbox{and}~~~~\nu >0.
\end{equation}
For this potential, Dirac equation can be written as
\begin{equation}
\left[\vec{\alpha}\cdot \vec{p}+\beta m+V(r)\right]
                 \Psi (r)=E\Psi (r)
\end{equation}
where $\vec{\alpha}$ and $\beta $ are Dirac matrixes.
Eq. (2) has two solutions with positive and negative energy
given respectively in the forms
\begin{equation}
\psi_\Lambda (r)=\left[
     \begin{array}{c}
           ig_\Lambda (r)/r \\
\vec{\sigma}\cdot\widehat{r}f_\Lambda (r)/r
\end{array}
\right] U_\Lambda (\widehat{r})
\end{equation}
\begin{equation}
\phi_\Lambda (r)=\left[
\begin{array}{c}
\vec{\sigma}\cdot\widehat{r}f_\Lambda (r)/r \\
g_\Lambda (r)/r
\end{array}
\right] \widetilde{U}(\widehat{r})
\end{equation}
where $\Lambda =(nljm)$ represents the set of Dirac quantum
numbers specifying the eigenmodes. The spin angular parts
$U_\Lambda (\widehat{r})$ and
$\widetilde{U}_\Lambda (\widehat{r})$ are described as
\begin{eqnarray}
U_{ljm}(\widehat{r})
&=&\sum_{m_l,m_s}\langle lm_l\frac 12m_s\left|jm
  \right\rangle Y_l^{m_l}(\widehat{r})\chi_{1/2}^{m_s},\\
\widetilde{U}_{ljm}(\widehat{r})
&=&(-1)^{j+m-l}U_{lj-m}(\widehat{r})
\end{eqnarray}
Substituting Eq. (3) or Eq. (4) into Eq. (2), one obtains
(for $n=0$, $l=0$)
\begin{equation}
\left[ \frac{-1}2\frac{d^2}{dr^2}+\lambda_qAr^\upsilon \right]
  g_\Lambda (r)=\lambda_q(E-m-V_0)g_\Lambda (r),
\end{equation}
\begin{equation}
f_\Lambda (r)=\frac 1{\lambda_q}(\frac d{dr}-
                       \frac 1r)g_\Lambda (r),
\end{equation}
where $\lambda_q=E+m$, $E$ is energy of confined
quark and $m$ is quark mass. Using the substitution
$\rho =(\lambda_qA)^{\frac 1{\nu +2}}r$ for convenience,
Eq. (7) reduces to the form
\begin{equation}
\left[ \frac{-1}2\frac{d^2}{d\rho ^2}+
         \rho ^\nu \right] g(\rho )=
               \epsilon g(\rho ),
\end{equation}
where
\begin{equation}
\epsilon =\left[ \frac{\lambda_q^\nu}{A^2}
       \right]^{\frac 1{\nu+2}}(E-m-V_0),
\end{equation}
and $g(\rho )$ is chosen as
\begin{equation}
g(\rho )\approx \rho \exp\left( -\left( x\rho \right) ^d\right),
\end{equation}
where $x$ and $d$ are variation parameters and they are obtained
by minimizing $\epsilon$. Hence they are solutions of
\begin{equation}
\frac{\partial \epsilon}{\partial x}=0,
\end{equation}
\begin{equation}
\frac{\partial \epsilon}{\partial d}=0.
\end{equation}
Using Eq. (9), Eq. (11) and Eq. (12), $\epsilon $ is found to be
\begin{equation}
\epsilon =\left( \frac{\nu +2}\nu \right) ax^2
\end{equation}
where $x=\left( \frac{b\nu}{2a}\right)^{\frac 1{\nu +2}}$,
 ~$a=\left( \frac{d+1}8\right)
     2^{\frac 2d} \frac{\Gamma
     \left( \frac 1d\right)}{\Gamma
     \left( \frac 3d\right)}$, ~$b=2^{\frac{-\nu}d}
     \frac{\Gamma \left(\frac{\nu +3} d\right)}
                   {\Gamma \left( \frac 3d\right)}$.
When Eq. (13) is calculated numerically, $d$ is obtained as
$1.502$ and using this value $\epsilon $ is found as $1.3268$
which is very close to the results of the WKB method and
$1/N$-expansion. Thus, the lowest eigenmodes corresponding
to the positive and negative energies have the respective
explicit forms
\begin{equation}
\psi_{\Lambda (1s_{1/2})}(r)=\frac 1{\sqrt{4\pi}}\left[
    \begin{array}{c}ig(r)/r \\
\vec{\sigma}\cdot\widehat{r}f(r)/r
\end{array}
\right] \chi_m,
\end{equation}
\begin{equation}
\phi_{\Lambda (1s_{1/2})}(r)=\frac 1{\sqrt{4\pi}}\left[
\begin{array}{c}
\vec{\sigma}\cdot\widehat{r}f(r)/r \\
-ig(r)/r
\end{array}
\right] \widetilde{\chi}_m,
\end{equation}
where the two component spinors $\chi_m$ and
$\widetilde{\chi}_m$ denote $ \chi_{\uparrow}=
\left(
\begin{array}{c}
1 \\
0
\end{array}
\right) $, $\chi_{\downarrow}=\left(
\begin{array}{c}
0 \\
1
\end{array}
\right) $ and $\widetilde{\chi}_{\uparrow}=\left(
\begin{array}{c}
0 \\
-i
\end{array}
\right) $, $\widetilde{\chi}_{\downarrow}=\left(
\begin{array}{c}
i \\
0
\end{array}\right) $. Using Eq. (8) and Eq. (5), the
radial parts in the upper and lower component solutions
corresponding to a quark flavor $q$ are
\begin{eqnarray}
g_q(r)&=&N_q\left( \frac r{r_{0q}}\right) \exp\left( -
      \left(\frac x{r_{0q}}\right) ^dr^d\right),\\
f_q(r)&=&-\frac{N_qd}{\lambda_qr_{0q}}\left(
   \frac x{r_{0q}}\right)^dr^d\exp\left( -\left(
        \frac x{r_{0q}}\right) ^dr^d\right),
\end{eqnarray}
where, $N_q$ is normalization constant obtained from the equation
\begin{eqnarray}
N_q^2\int_0^\infty \left( f_q^2(r)+g_q^2(r)\right) dr=1.
\end{eqnarray}

\section{Quark-antiquark momentum distribution}

Knowing the quark-antiquark eigenmodes in the ground-state of
meson, it is possible to obtain their corresponding momentum
distribution amplitude. If $G_q(p,\lambda ,\lambda ^{\prime})$
is the amplitude for finding a bound quark of flavor $q$ in its
eigenmode $\Phi_{q\lambda}^{\left( +\right)}(r)$ in a state
of definite momentum $p$ and spin projection $\lambda ^{\prime}$,
then it is given by$^{15,17}$
\begin{eqnarray}
\Phi_{q\lambda}^{\left( +\right)}(r)=
  \sum_{\lambda ^{^{\prime}}}\int d^3pG_q(p,\lambda,
  \lambda^{\prime})\sqrt{\frac m{E_p}}
  U_q(p,\lambda^{\prime})\exp \left( i\vec{p}\cdot\vec{r}\right).
\end{eqnarray}
Where $U_q(p,\lambda ^{\prime})$ is the usual free
Dirac spinors. Eq. (20) can be easily inverted to yield
\begin{equation}
G_q(p,\lambda ,\lambda ^{\prime})=
  \sqrt{\frac m{E_p}}U_q^{\dagger}(p,\lambda ^{^{\prime}})
  \int d^3r\Phi_{q\lambda}^{\left( +\right)}(r)
  \exp\left( -i\vec{p}\cdot\vec{r}\right).
\end{equation}
Thus, it is found as
\begin{equation}
G_q(p,\lambda ,\lambda ^{\prime})=
    G_q(p) \delta_{\lambda \lambda ^{\prime}},
\end{equation}
where
\begin{equation}
G_q(p)=\frac{i\pi N_q}{\sqrt{2}dr_{0q}
       \beta ^{\frac 3d}\lambda_q}(E_p+E_q)
       \sqrt{\frac{m+E_p}{E_p}}H(d,z),
\end{equation}
where, $E_p=\sqrt{p^2+m^2}$, $E_q$ is the solution of Eq. (10),
$\beta=\left( \frac x{r_{0q}}\right) ^d$,
$z=\frac{pr_{0q}}{2x}$ and
\begin{eqnarray}
H(d,z)=\frac{d^2}{2^{\frac{2d+3}{2}}}z^{\frac{-3}{2}}
  \int_0^\infty y^{\frac{2d+1}2}\exp\left( -
  \left( \frac y2\right) ^d\right) J_{3/2}(yz)dy.
\end{eqnarray}

\section{Leptonic decay widths}

Now following Margolis and Mendel$^{17}$ one can represent
the ground state of a neutral vector meson such as
($\rho $, $\omega $, $\phi $) with a
particular spin projection $S_V$ and zero momentum as
\begin{equation}
\left| V(0),S_V\right\rangle =
  \frac{\sqrt{3}}{\sqrt{N(0)}}\sum_{q,\lambda_1,
  \lambda_2}\int d^3pG_q(p)C_{\lambda_1\lambda_2}^{S_V}
  \zeta_q^Vb_q^{\dagger}(p,\lambda_1)\widetilde{b}_q^{\dagger}
  (-p,\lambda_2)\left| 0\right\rangle.
\end{equation}
Here, $b_q^{\dagger}(p,\lambda_1)$ and
$\widetilde{b}_q^{\dagger}(-p,\lambda_2)$ operating on
the vacuum state are quark and antiquark creation operators,
respectively. The summation with the flavor coefficient
$\zeta_q^V$ and the spin configuration coefficient
$C_{\lambda_1\lambda_2}^{S_V}$ represents the appropriate
SU(6) spin-flavor structure of the particular vector meson
$V$ with its spin projection $S_V$ and zero momentum. The factor
$\sqrt{3}$ is due to the effective color singlet configuration
of the meson. $N(0)$ represents the overall normalization,
which is given by
\begin{eqnarray}
N(0)=\frac 1{(2\pi )^3}\int_0^\infty d^3p\left| G_q(p)\right| ^2.
\end{eqnarray}
Assuming that the main contribution to the leptonic decay process
of neutral vector mesons such as $\rho $, $\omega $, $\phi $
comes from single virtual photon creation from the annihilation
of the bound quark-antiquark pair inside of the meson, $S$-matrix
element in configuration space can be written as$^{15}$
\begin{eqnarray}
S_{fi}&=&\left\langle e^{-}(k_1,\delta_1)e^{+}(k_2,\delta_2)
  \right|-ie^2\int d^4x_1d^4x_2(\overline{\psi}_e(x_2)
  \gamma^\mu \psi_e(x_2)D_{\mu\nu}(x_2-x_1)  \nonumber \\
&\times &\sum_qe_q\overline{\psi}_q(x_1)\gamma ^\nu \psi_q(x_1))
 \left|V,S_V\right\rangle
\end{eqnarray}
Where, $D_{\mu \nu}(x_2-x_1)$ is the photon propagator,
$\psi_e(x)$ and $\psi_q(x)$ are the free lepton and quark
fields, respectively. After some standard calculations (details
of the calculation can be found in Ref. 15), one obtains
\begin{equation}
\Gamma \left( V\rightarrow e^{+}e^{-}\right) =\frac{4\pi}3\alpha
_{em}^2M_Vf_V^2.
\end{equation}
Where, $f_V$ is known as the leptonic decay constant
and in this model it
can be written as
\begin{equation}
f_V^2=\frac{2\left\langle e_q\right\rangle_V^2I_V^2}
                    {3\pi^2M_V^3J_V}.
\end{equation}
Here,
\begin{eqnarray}
\left\langle e_q\right\rangle_{\rho ,\omega ,\phi}
&=&\left( \frac 1{\sqrt{2}},\frac 1{3\sqrt{2}},
      \frac 13\right), \nonumber \\
I_V&=&\int_{0}^{\infty} dpp^2\left( 2+\frac m{E_p}\right) G_q(p),
                                    \nonumber \\
J_V&=&\int_0^\infty dpp^2\left| G_q(p)\right|^2.
\end{eqnarray}
$I_V$ and $J_{V\;}$values can be calculated numerically by computer.

\section{Weak leptonic decay widths}

Here, the weak leptonic decay of charged pseudoscalar mesons such
as $\pi ^{\pm}$, $K^{\pm}$, $D^{\pm}$, $D_s^{\pm}$, $B^{\pm} $
and $B_c^{\pm}$ are considered. Assuming that the main
contribution to the weak leptonic decay processes comes from the
single virtual boson creation from the annihilation of the bound
quark-antiquark pair inside the pseudoscalar meson $M$, the
$S$-matrix element in configuration space is written as$^{16}$
\begin{eqnarray}
S_{fi} &=&\left\langle l(k_1,\delta_1)\overline{\nu}_l(k_2,
     \delta_2)\right| \left( \frac{-iG_F}{\sqrt{2}}\right)
     \int d^4x\overline{\psi}_l(x)\gamma ^\mu
     \left(1-\gamma^5\right) \psi_l(x)  \nonumber \\
&\times &\sum_{qm,qn}\nu_{q_mq_n}\psi_{q_m}(x)\gamma_\mu
    \left( 1-\gamma^5\right) \psi_{q_n}(x)\left| M(0)\right\rangle.
\end{eqnarray}
Where, $G_F$ is the Fermi coupling constant, $\nu_{q_mq_n}$
are the CKM matrix elements and $\left| M(0)\right\rangle $
is given by$^{16}$
\begin{eqnarray}
\left| M(0)\right\rangle
&=&\sqrt{\frac 3{N(0)}}\sum C_{q_1q_2}^M
             (\lambda_{1,}\lambda_2) \nonumber\\
&\times&\int d^3p\left[ G_{q_1}(p)G_{q_2}^{*}
  (-p)\right]^{\frac12}b_{q_1}^{\dagger}(p,\lambda_1)
    \widetilde{b}_{q_2}^{\dagger}
      (-p,\lambda_2)\left| 0\right\rangle,
\end{eqnarray}
where, $C_{q_1q_2}^M(\lambda_{1,}\lambda_2)$ stands for the
appropriate SU(6) spin-flavor coefficients for the pseudoscalar
meson $M$. $N(0)$ represents the overall normalization, which
is given by
\begin{equation}
N(0)=\frac 1{(2\pi )^3}\int_0^\infty d^3p \left(
G_{q_1}(p)G_{q_2}^{\star}(-p)\right).
\end{equation}
After some standard calculations, which can be found
in Ref. 16, one obtains
\begin{equation}
\Gamma \left( M\rightarrow l\overline{\nu}_l\right) =
 \frac{G_F^2}{8\pi}\left| \nu_{q_1q_2}\right|^2M_pm_l^2
 \left( 1-\frac{m_l^2}{M_p^2}\right)^2f_M^2,
\end{equation}
where $f_M$ is the weak decay constant, having the form
\begin{equation}
f_M^2=\frac{3I_M^2}{2\pi ^2M_pJ_M}.
\end{equation}
Here $M_p$ is the mass of pseudoscalar meson, $m_l$ is the
mass of lepton and $I_M$ and $J_M$ are found as
\begin{eqnarray}
I_M&=&\int_0^\infty dpp^2A(p)\left[ G_{q_1}(p)
   G_{q_2}^{*}(-p)\right] ^{\frac12}, \nonumber \\
J_M&=&\int_{0}^{\infty} dpp^2\left[ G_{q_1}(p)
                          G_{q_2}^{*}(-p)\right],
\end{eqnarray}
and
\begin{equation}
A(p)=\frac{\left( E_{p_1}+m_{q_1}\right) \left( E_{p_2}+
    m_{q_2}\right) -p^2}{\left[ E_{p_1}E_{p_2}\left( E_{p_1}+
    m_{q_1}\right) \left(E_{p_2}+m_{q_2}\right)
                        \right] ^{\frac 12}},
\end{equation}
where $E_{p_i}=\sqrt{p^2+m_{q_i}^2}$.

\section{Results}

The calculations involve the potential parameters of the model
$\left(A,V_0,\nu \right) $ and the quark masses
$\left( m_u=m_d,m_s,m_c,m_b\right)$ .
The potential parameters are chosen to be
$A=0.68 \mbox{GeV},~V_0=-0.3961 \mbox{GeV},~\nu = 0.2$.

The light-quark masses $m_u=m_d$ and $m_s$ are obtained from
$\omega $ and $\phi $ mesons as
$m_u=m_d=0.078 \mbox{GeV}$, $m_s = 0.3 \mbox{GeV}$,
and from $D^{\pm}$ and $B^{\pm}$
$m_c=1.3 \mbox{GeV}$, $m_b =4.81 \mbox{GeV}$.

When these parameters are used the masses of
$D,D_s,B,B_s,B_c,w$ and $\phi $ can be calculated
to be almost the same with their experimental values.
Also, with the same parameters, radiative decay widths
of light and heavy mesons have been calculated and the
obtained results are close to their experimental
values$^{8}$. Using this potential parameters and quark
masses given above, the calculated results are shown in
Table I, Table II and Table III.

\section{Conclusion}

In this paper, the independent particle model approach has
been used to investigate leptonic decay of light vector mesons
and weak leptonic decay of light and heavy pseudoscalar mesons.
The Dirac equation in a power-law potential has been solved with
variation technique using special type wave function. Leptonic
decay widths and decay constant $f_V$ and weak leptonic decay
widths and decay constant $f_M$ have been calculated and are
compared with the results of other theoretical investigations
and experiments. The results say to us that the potential used
in this study can be considered as interaction potential for
quarks. Because of the model structure it is assumed that quarks
do not interact with each other. However, there are spin-spin and
other hyperfine interactions between quarks. Adding these
interactions, good results, especially in the calculation of
meson mass spectra, could be obtained. Such interaction terms
can be found in Ref. 24,25.

\section*{Acknowledgments}

We thank T.M. Aliev and H. Akcay for useful discussions.

\section*{References}

\begin{itemize}
\begin{enumerate}
\item  N.Barik, B.K.Dash and M.Das, {\it Phys.Rev.}
                          {\bf D31} (1985) 1652,\\
       P.Leal Ferrera, {\it Lett. Nuovo cimento}
                                 {\bf 20} (1977) 157,\\
       P.Leal Ferrera and N.Zagury, {\it ibid} {\bf 20} (1977) 511
\item  N.Barik and B.K.Dash, {\it Phys. Rev.} {\bf D33} (1986) 1925
\item  N.Barik and B.K.Dash, {\it Phys. Rev.} {\bf D34} (1986) 2803
\item  N.Barik, B.K.Dash and M.Das, {\it Phys. Rev.}
                                {\bf D32} (1985) 1725
\item  N.Barik and B.K.Dash, {\it Phys. Rev.} {\bf D34} (1986) 2052
\item  H.Akcay and H.Ciftci, {\it J. Phys.}  {\bf G22} (1996) 455
\item  N.Barik, B.K.Dash and P.C.Dash, {\it Pramana J. Phys.}
               {\bf 29} (1987) 543
\item  N.Barik, P.C.Dash and A.R.Panda, {\it Phys. Rev.}
            {\bf D46} (1992) 3856,\\
       N.Barik, P.C.Dash, {\it Phys.Rev.} {\bf D49} (1994) 299,
       H.\d{C}iftci (unpublished)
\item  A.W.Thomas, {\it Adv. Nucl. Phys.} {\bf 13} (1983) 1
\item  R.Van Royen and V.F.Weisskopf, {\it Nuovo cimento}
                             {\bf A50} (1967) 617
\item  S.N.Sinha, {\it Phys. Lett.} {\bf B178} (1986) 110,\\
       G.Godfrey, {\it Phys. Rev.} {\bf D33} (1986) 1391
\item  L.J.Reinders, {\it Phys. Rev.} {\bf D38} (1988) 947,\\
       S.Narison, {\it Phys. Lett.} {\bf B198} (1987) 104,\\
       C.A.Dominguez and N.Power, {\it Phys. Lett.} {\bf B197}
             (1987) 423
\item  M.B.Govale, {\it et al., Phys. Lett.}
                                     {\bf B206} (1988) 113,\\
       R.M.Wokshyn {\it et al. Phys. Rev.} {\bf D39} (1989) 978
\item  S.Capstick and S.Godfrey, {\it Phys. Rev.}
                          {\bf D41} (1990) 2856
\item  N.Barik, P.C.Dash and A.R.Panda, {\it Phys. Rev.}
       {\bf D47} (1993) 1001
\item  N.Barik and P.C.Dash, {\it Phys. Rev.}
                                         {\bf D47} (1993) 2788
\item  B.Margolis and R.R.Mandal, {\it Phys. Rev.}
                                        {\bf D28} (1983) 468,\\
       C.Hayne and N.Isgnur, {\it Phys.Rev.}
                                          {\bf D25} (1982) 1944
\item  H.Kraseman, {\it Phys.Lett.} {\bf B96} (1980) 397
\item  E.Glowich, {\it Phys.Lett.} {\bf B91} (1980) 271
\item  M.Claudsun, {\it Harward Report No.}
                              {\bf 91} (1981) (unpublished)
\item  C.Bernard, {\it et al., Phys. Rev.}
                                          {\bf D38} (1980) 3540
\item  H.W.Hamber, {\it Phys. Rev.} {\bf D39} (1989) 896
\item  C.Caso {\it et al., The European Physical Journal}
       {\bf C3} (1998) 1
\item  Ho-Meoyng Choi and Chueng-Ryong Ji, {\it Phys. Lett.}
                     {\bf B460} (1999) 461
\item  Ho-Meoyng Choi and Chueng-Ryong Ji, {\it Phys. Rev.}
                               {\bf D59} (1999) 074015
\end{enumerate}
\end{itemize}

\newpage
Table I. Leptonic decay widhts and the decay constant
$f_V$ in keV in comparison with the results of other
researchers and the experiment.

\begin{center}
\begin{tabular}{c|ccccc}\hline\hline
Meson & $\Gamma \left( V\rightarrow e^{+}e^{-}\right)$
     & Experiment$^{23}$ &Ref. 15 & Ref. 17 &  \\ \hline
$\rho $ & 6.37 & 6.77$\pm $0.32 & 6.26(8.1) & 7.8 &  \\
$\omega $ & 0.684 & 0.6$\pm $0.02 & 0.67(0.87) & 0.84 &  \\
$\phi $ & 1.46 & 1.37$\pm $0.05 & 1.58(1.84) & 1.69 &  \\
 & & & & &  \\
                                               \hline \hline
Meson & $f_V$ & Experiment$^{23}$ & Ref. 15 & Ref. 17 &  \\
                                                 \hline
$\rho $ & 0.193 & 0.2$\pm $0.04 & 0.19(0.22) & 0.21 &  \\
$\omega $ & 0.0624 & 0.06$\pm $0.01 & 0.06(0.07) & 0.07 &  \\
$\phi $ & 0.0813 & 0.08$\pm $0.01 & 0.08(0.09) & 0.07 &  \\ \hline
\end{tabular}
\end{center}

\newpage
Table II. Decay constants of pseudoscalar mesons in MeV
in comparison with the results of other model and the
experiment. Experimental values are taken from Ref. 16.

\begin{center}
\begin{tabular}{c|cccccccc}\hline\hline
Model & $f_\pi $ & $f_K$ & $f_D$ & $f_{D_s}$ & $f_B$
                       & $f_{B_s}$ & $f_{B_c}$ &  \\ \hline
Expt.$^{16,24}$ & 131.73$\pm $0.15 & 160.6$\pm $1.3
                       & $<$219 & 137-304 & - & -& - &  \\
This work & 131.4 & 150.2 & 181.1 & 235.9 & 208.3
                          & 257.3 & 377.2 &  \\
Ref. 16 & 138 & 157 & 161 & 205 & 122 & 154 & 221 &  \\
Ref. 14 & 100 & 153 & 149 & 160 & 96 & 111 & 141 &  \\
Ref. 20 & - & - & 172 & 196 & 149 & 170 & 255 &  \\
Ref. 18 & 139 & 176 & 150 & 210 & 125 & 175 & 425 &  \\
Ref. 19 & 178 & 182 & 148 & 166 & 98 & - & - &  \\
Ref. 21 & - & - & 174$\pm $53 & 234$\pm $72
                   & 105$\pm $34 & 155$\pm $75 & -&  \\
Ref. 22 & 141$\pm $21 & 155$\pm $21 & 282$\pm $28
                  & - & 183$\pm $28 & - & -&  \\ \hline
\end{tabular}
\end{center}

\newpage
Table III. Partial decay widths
$\Gamma \left( M\rightarrow l\overline{\nu_l}\right)$
in MeV and the branching ratio
$B\left( M\rightarrow l\overline{\nu_l}\right)$ of
pseudoscalar mesons in comparison with the experiment.\.
($B\left( M\rightarrow l\overline{\nu_l}\right) =
\tau_M\Gamma\left(M\rightarrow l\overline{\nu_l}\right) $).
$\tau_M$ is the mean lifetime of meson $M$

\begin{center}
\begin{tabular}{c|ccc}\hline\hline
Prosses & $\Gamma \left( M\rightarrow l\overline{\nu_l}\right)
   $ & $B\left(M\rightarrow l\overline{\nu_l}\right) $
   & Expt.$^{23}$ $B\left( M\rightarrow l\overline{\nu_l}
          \right) $ \\ \hline
$\pi ^{\pm}\rightarrow \mu ^{\pm}\overline{\nu}_\mu $
  & 2.523 10$^{-14}$ & 0.994 & 0.999877$\pm $0.0000004 \\
$\pi ^{\pm}\rightarrow e^{\pm}\overline{\nu}_e$
  & 3.238 10$^{-18}$ & 1.276 10$^{-4}$
  & (1.23$\pm $0.004) 10$^{-4}$ \\
$K^{\pm}\rightarrow \mu ^{\pm}\overline{\nu}_\mu $
  & 3.000 10$^{-14}$ & 0.563 & 0.6351$\pm $0.0018 \\
$K^{\pm}\rightarrow e^{\pm}\overline{\nu}_e$
  & 7.724 10$^{-19}$ &1.45 10$^{-5}$
           & (1.55$\pm $0.07) 10$^{-5}$ \\
$D^{\pm}\rightarrow \mu ^{\pm}\overline{\nu}_\mu $
  & 1.795 10$^{-13}$ & 2.87 10$^{-4}$
  & $<$7.2 10$^{-4}$ \\
$D_s^{\pm}\rightarrow \mu ^{\pm}\overline{\nu}_\mu $
  & 6.240 10$^{-12}$ & 4.41 10$^{-3}$
  & 4$_{-2.0}^{+2.2}\times $10$^{-3}$ \\
$B^{\pm}\rightarrow \mu ^{\pm}\overline{\nu}_\mu $
  & 1.693 10$^{-16}$ & 4.15 10$^{-7}$
  & $<$2.1 10$^{-5}$ \\
$B_c^{\pm}\rightarrow \mu ^{\pm}\overline{\nu}_\mu $
  & 8.960 10$^{-14}$ & - & - \\
$D^{\pm}\rightarrow \tau ^{\pm}\overline{\nu}_\tau $
  & 4.720 10$^{-13}$ & 7.540 10$^{-4}$ & - \\
$D_s^{\pm}\rightarrow \tau ^{\pm}\overline{\nu}_\tau $
  & 6.090 10$^{-11}$ & 4.3 10$^{-2}$
  & (7$\pm $4) 10$^{-2}$ \\
$B^{\pm}\rightarrow \tau ^{\pm}\overline{\nu}_\tau $
  & 3.770 10$^{-14}$ & 9.25 10$^{-5}$
  & $<$0.57 10$^{-3}$ \\
$B_c^{\pm}\rightarrow \tau ^{\pm}\overline{\nu}_\tau $
  & 2.140 10$^{-11}$ & - & - \\ \hline
\end{tabular}
\end{center}

\end{document}